\documentclass[submission,copyright,creativecommons]{eptcs}
\providecommand{\event}{HCVS 2022} 
\usepackage{breakurl}             
\usepackage{multirow}
\usepackage{amssymb}
\usepackage{xspace}
\usepackage[noabbrev,capitalize]{cleveref}
\usepackage{hhline}
\usepackage[htt]{hyphenat}
\usepackage{wrapfig}
\title{CHC-COMP 2022: Competition Report}
\author{
Emanuele De Angelis\thanks{
The author is member of the INdAM Research Group GNCS.
}
\institute{IASI-CNR, Italy}
\email{emanuele.deangelis@iasi.cnr.it}
\and
Hari Govind V K
\institute{University of Waterloo, Canada}
\email{hgvk94@gmail.com}
}
\def\titlerunning{CHC-COMP 2022: Competition Report}
\def\authorrunning{Emanuele De Angelis and Hari Govind V K}
\newcommand{\spacer}{Spacer}
\newcommand{\eldarica}{Eldarica}
\newcommand{\golem}{Golem}
\newcommand{\unihorn}{U. Unihorn}
\newcommand{\treeautomizer}{U. TreeAutomizer}
\newcommand{\ringen}{RInGen}
\usepackage[table,x11names]{xcolor}

\begin{document}
\maketitle
\begin{abstract}
CHC-COMP 2022 is the fifth edition of the competition
of solvers for Constrained Horn Clauses.
The competition was run in March 2022; the results
were presented at the 9th Workshop on Horn
Clauses for Verification and Synthesis held in Munich,
Germany, on April 3, 2022.
This edition featured six solvers, and eight tracks
consisting of sets of linear and nonlinear clauses with
constraints over linear integer arithmetic, linear real
arithmetic, arrays, and algebraic data types.
This report provides an overview of the organization
behind the competition runs: it includes the technical
details of the competition setup as well as presenting
the results of the 2022 edition.
\end{abstract}
\section{Introduction}
\textit{Constrained Horn Clauses} (CHCs, for short) are
a class of first-order logic formulas where the Horn clause 
format is extended with \textit{constraints}, that is,
formulas of an arbitrary, possibly non-Horn, background
theory (such as linear integer arithmetic, arrays, and
algebraic data types).

CHCs have been advocated by many researchers as suitable  
framework to model software systems and to reason about
their properties 
(see, for instance,~\cite{BGMR2015,DFGHPP2021,Gurfinkel2022}).
In particular, CHCs turn out to be a very flexible and 
expressive intermediate language to formalize features of
different programming and specification languages as well
as proof systems used in the development of software systems,
thereby making solvers for CHCs  (CHC solvers, for short) 
complementary tools that can be conveniently used to 
automate the analysis and verification pipeline of software
systems.

CHC-COMP 2022\footnote{The CHC-COMP 2022 webpage is 
available at~\url{https://chc-comp.github.io/}} is the
5th edition of the competition of solvers for CHCs, 
affiliated with the 9th Workshop on Horn Clauses for
Verification and
Synthesis (HCVS 2022\footnote{The HCVS 2022 webpage is 
available at \url{https://www.sci.unich.it/hcvs22/}.}) 
held in Munich, Germany, on April 3, 2022.
The goal of the CHC-COMP series is to evaluate the 
effectiveness and the efficiency of state-of-the-art 
solvers for CHCs on realistic
and publicly available benchmarks.

CHC-COMP is open to proposals for new competition tracks
and anyone is welcome to submit candidate benchmarks.
The CHC-COMP 2022 deadline for submitting benchmarks to be
considered for the competition was March 1, 2022. 
The deadline for submitting the solvers for test runs 
(optional, but recommended) was March 8, 2022, 
while the deadline for submitting the solvers for evaluation 
was March 18, 2022 (extended to March 21, 2022).
The competition was run in the subsequent two weeks,
and the results were announced at the HCVS workshop
on April 3, 2022.

CHC-COMP 2022 featured 6 solvers, and 8 tracks consisting
of sets of linear and nonlinear clauses with constraints
over linear integer arithmetic, linear real arithmetic, 
arrays, algebraic data types, and a few combinations of
such theories.

This report is structured as follows. 
Section~\ref{sec:D&O} presents the competition tracks,
the technical resources used to run the competition,
and the evaluation model adopted to rank the solvers.
Section~\ref{sec:benchmarks} presents the benchmark 
suite, specifically, the format, the inventory, and
how the candidate benchmarks have been processed and
selected for the competition runs.
Section~\ref{sec:solvers} presents the tools that were
submitted to CHC-COMP 2022.
Section~\ref{sec:results} presents the results of this 
edition. 
Section~\ref{sec:conclusions} presents a few concluding
remarks from the organizers, which also include the 
suggestions and observations from the participants.
Section~\ref{sec:tools} collects the tool 
descriptions contributed by the participants.
Finally, Appendix~\ref{app:detres} includes the tables 
with the detailed results about the competition runs.

\subsection*{Acknowledgements} 
We would like to thank the HCVS 2022 Program Chairs,
Temesghen Kahsai and Maurizio Proietti, for hosting
the competition this year as well.

CHC-COMP 2022 heavily built on the infrastructure developed
by the Organizers of the previous editions, that is, 
Grigory Fedyukovich, Arie Gurfinkel, and Philipp R\"{u}mmer,
which also includes the contributions from Nikolaj Bj{\o}rner, 
Adrien Champion, and Dejan Jovanovic.

We are deeply indebted to StarExec~\cite{starexec} that 
continues to provide the CHC-COMP community the computing 
resources and evaluation environment to run the competition. 
In particular, we would like to thank Aaron Stump for his 
indispensable support with StarExec. 

We are also extremely grateful to Philipp R\"{u}mmer who 
kindly helped us by answering our numerous questions about
the design and the organization of the competition.

Finally, we would to thank all the CHC-COMP 2022 participants
for having submitted their solvers, as well as all who
contributed with new benchmarks, and all the HCVS 2022 
participants for the profitable discussion we had at the 
workshop after the presentation of the results.

\section{Design and Organization}
\label{sec:D&O}

This section presents 
(i) the competition tracks,
(ii) the technical resources used to run the solvers, and
(iii) the evaluation system used to rank the solvers in
each track.

\newcommand{\LIAlin}{LIA-lin\xspace}
\newcommand{\LIAnonlin}{LIA-nonlin\xspace}
\newcommand{\LIAarrlin}{LIA-lin-Arrays\xspace}
\newcommand{\LIAarrnonlin}{LIA-nonlin-Arrays\xspace}
\newcommand{\LRATSlin}{LRA-TS\xspace}
\newcommand{\LRATSpar}{LRA-TS-par\xspace}
\newcommand{\ADTnonlin}{ADT-nonlin\xspace}
\newcommand{\ADTarrnonlin}{LIA-nonlin-Arrays-nonrecADT\xspace}

\subsection{Tracks}
Solvers participating in the CHC-COMP 2022 could 
enter the competition in eight tracks: one track was
introduced in this edition, that is, \ADTarrnonlin,
while the remaining tracks were inherited from the 
previous editions. 

The tracks are classified according to type of clauses 
included in the corresponding benchmarks.
In particular, the categories have been defined by 
considering the following features:
(i) the background theory of the constraints, and 
(ii) the structure of the clauses, that is, linear clauses 
(clauses with at most one uninterpreted atom in the premise
of the clause), and nonlinear clauses (clauses with more
than one uninterpreted atom in the premise of the clause).

We have considered the following tracks, which includes
benchmarks of linear and nonlinear clauses combining 
various background theories.

\begin{itemize}

\item \textbf{\LIAlin}:
Linear Integer Arithmetic -- linear clauses.

\item \textbf{\LIAnonlin}:
Linear Integer Arithmetic -- nonlinear clauses.

\item \textbf{\LIAarrlin}:
Linear Integer Arithmetic and Arrays -- linear clauses.

\item \textbf{\LIAarrnonlin}:
Linear Integer Arithmetic and Arrays -- nonlinear clauses. 

\item \textbf{\ADTnonlin}:
Algebraic Data Types -- nonlinear clauses.

\item \textbf{\ADTarrnonlin}:
Linear Integer Arithmetic, Arrays and nonrecursive 
Algebraic Data Types -- nonlinear clauses.

\end{itemize}

Moreover, we have also considered two additional tracks
including a syntactically restricted class of linear 
clauses, that is, transition systems.
Benchmarks in this track have exactly one uninterpreted 
relation symbol, and exactly three linear clauses encoding
initial states, transitions, and error states.

\begin{itemize}

\item \textbf{\LRATSlin}: Linear Real Arithmetic --
Transition Systems.

\item \textbf{\LRATSpar}: Linear Real Arithmetic --
Transition Systems -- parallel.

\end{itemize}

\subsection{Technical Resources}

CHC-COMP 2022 was run on StarExec 
(\url{https://www.starexec.org/})
using the same technical resources used in the 2021
edition~\cite{chccomp21}.
For the sake of completeness, we summarize the details
of technical resources available to run the competition.

StarExec made available to CHC-COMP 2022 two queues:
\texttt{chcpar.q} and \texttt{chcseq.q}, consisting
of 15 and 20 nodes, respectively, each of which equipped
with two quadcore CPUs. The machine specifications (see
\url{https://www.starexec.org/starexec/public/machine-specs.txt}) are:

\smallskip

\begin{verbatim}
# Starexec stats nodes 001 - 192:
Intel(R) Xeon(R) CPU E5-2609 0 @ 2.40GHz (2393 MHZ)
    10240  KB Cache
    263932744 kB main memory

# Software:
OS:       CentOS Linux release 7.7.1908 (Core)
kernel:   3.10.0-1062.4.3.el7.x86_64
glibc:    glibc-2.17-292.el7.x86_64
          gcc-4.8.5-39.el7.x86_64
          glibc-2.17-292.el7.i686
\end{verbatim}

\medskip

Running a solver on a track was performed by submitting
a job to a StarExec node. A job is a pair consisting of
a solver (with a track-specific configuration) and a
benchmark.

Each job of LRA-TS-par was run on a node of the 
\texttt{chcpar.q} queue, while two jobs of all other
tracks were run in parallel on a node of the the
\texttt{chcseq.q} queue.

\subsection{Test and Competition Runs}

Similar to CHC-COMP 2021~\cite{chccomp21}, the submitted
solvers were evaluated twice by performing a \textbf{test}
run and a \textbf{competition} run.
For the sake of completeness, in the following we summarize
the main features of the two kinds of runs (see also the 
report~\cite{chccomp21} of the 2021 edition). 

In the \textbf{test} runs, the (optional) pre-submissions of the
solvers were evaluated to check their configurations and
identify possible inconsistencies.
In these tests a small set of randomly selected benchmarks
was used, and each job was limited to 600s CPU time, 
600s wall-clock time, and 64GB memory.

In the \textbf{competition} runs, the final submissions
of the
solvers were evaluated to determine the outcome of the
competition, that is, to rank the solvers entering 
CHC-COMP 2022. The ranking method is presented in 
Section~\ref{subsub:eval}, while the process for 
selecting the benchmarks is described in 
Section~\ref{subsub:selBench}.
In the competition runs of LRA-TS-par each job was 
limited to 1800s wall-clock time, and 64GB memory.
For all other tracks, each job was limited to 1800s 
CPU time, 1800s wall-clock time, and 64GB memory.

\subsection{Evaluation of the Competition Runs}
\label{subsub:eval}

The competing solvers were evaluated using the same
approach as the 2021 edition~\cite{chccomp21}.

The evaluation of the competition runs were done using
the \texttt{summarize.py} script available at 
\url{https://github.com/chc-comp/scripts}; the script
takes as input the `job information' CSV file produced 
by StarExec at job completion.

The ranking of solvers in each track is based on the
\textbf{Score} obtained by the solvers in the competition
run for a track. The \textbf{Score} is computed on the
basis of the \textbf{result} provided by the solver on
the benchmarks for that track.
The \textbf{result} can be \textit{sat}, \textit{unsat},
or \textit{unknown} (which includes solvers giving up, 
running out of resources, or crashing), and the 
\textbf{Score} is given by the number of \textit{sat} or
\textit{unsat} results.

If two solvers reached an equal \textbf{Score},
the ranking was determined by using the \textbf{CPU time}
for all tracks except the LRA-TS-par track, where the
\textbf{Wall-clock time} is used instead of the 
\textbf{CPU time}.
The \textbf{CPU time} is the total CPU time needed by a
solver to produce a \textbf{result} in some track;
the \textbf{Wall-clock time} is the total 
wall-clock time needed by a solver to produce its answers
in some track.

The tables in Appendix~\ref{app:detres} also report in 
column `\#unique' the number of \textit{sat} or 
\textit{unsat} results produced by a solver for benchmarks
for which all other solvers returned \textit{unknown}.
The `job information' files also include data
about the space and memory consumption, which we consider
less relevant and therefore are not reported in the tables
(see also the CHC-COMP 2021 report~\cite{chccomp21}).

\section{Benchmarks}
\label{sec:benchmarks}

\subsection{Format}
CHC-COMP accepts benchmarks in the SMT-LIB 2.6 
format~\cite{BarFT-SMTLIB}.
All benchmarks have to conform to the format described
at \url{https://chc-comp.github.io/format.html}.
Conformance is checked using the \texttt{format.py} script
available at \url{https://github.com/chc-comp/scripts}.

\subsection{Inventory}
All benchmarks used for the competition are selected 
from repositories under \url{https://github.com/chc-comp}.
Anyone can contribute benchmarks to this repository.
This year, we got several new benchmarks for many of 
the tracks. Table~\ref{tab:bench_summary} summarizes 
the number of benchmarks and unique benchmarks available
in each repository. The organizers pick a subset of all
available benchmarks for each year's competition.
In the rest of this section, we explain the steps in
this selection.

\subsection{Processing Benchmarks without Algebraic Data Types or Reals}
All benchmarks are processed using the \texttt{format.py}
script, which is available at 
\url{https://github.com/chc-comp/scripts}.
The command line for invoking the script is

\noindent
\begin{verbatim}
   > python3.9 format.py --out-dir <out-dir> --merge_queries True <smt-file> 
\end{verbatim}

\noindent
The script attempts to put benchmark \texttt{<smt-file>} into CHC-COMP format.
The \texttt{merge\_queries} option merges multiple queries into a single query as discussed in previous editions of CHC-COMP~\cite{chccomp21}.

After processing, benchmarks are categorized into one of  4 competition tracks: LIA-lin, LIA-nonlin, LIA-lin-Arrays, and LIA-nonlin-Arrays. The scripts for categorizing the benchmarks are available at \url{https://github.com/chc-comp/chc-tools}. Benchmarks that could not be put in CHC-COMP compliant format and benchmarks that could not be categorized into any tracks are not used for the competition.

\bigskip
\begin{table}[!ht]
\centering
\begin{tabular}{
@{\hspace{0pt}}l@{\hspace{0pt}}
@{\hspace{4pt}}c@{\hspace{1pt}}
@{\hspace{5pt}}c@{\hspace{1pt}}
@{\hspace{5pt}}c@{\hspace{1pt}}
@{\hspace{1pt}}c@{\hspace{1pt}}
@{\hspace{0pt}}c@{\hspace{1pt}}
@{\hspace{0pt}}c@{\hspace{1pt}}
@{\hspace{0pt}}c@{\hspace{1pt}}}
\hline
Repository 
&  \multirow{1}{1.6cm}{\LIAlin}     
&  \multirow{2}{1.6cm}{\LIAnonlin}   
&  \multirow{2}{1.6cm}{\LIAarrlin} 
&  \multirow{2}{1.6cm}{\LIAarrnonlin} 
&  \multirow{1}{1.6cm}{\LRATSlin}
&  \multirow{2}{1.5cm}{\ADTnonlin}
&  \multirow{4}{1.8cm}{\ADTarrnonlin}\\
                     &               &            &            &         & & &\\
                     &               &            &            &         & & &\\
                     &               &            &            &         & & &\\\hline\hline
adt-purified         &               &            &            &         & & 67/67 &\\\hline
aeval                &  54/54        &            &            &         & & &\\\hline
aeval-unsafe~(\textit{new}) &  54/54 &            &            &         & & &\\\hline
chc-comp19           &               &            & 290/290    &         & 228/226 & &\\\hline
eldarica-misc        &  149/136      &  69/66     &            &         & & &\\\hline
extra-small-lia      &  55/55        &            &            &         & & &\\\hline
hcai                 &  101/87       &  133/131   & 39/39      &   25/25 & & &\\\hline
hopv                 &  49/48        &   68/67    &            &         & & &\\\hline
jayhorn              &  75/73        & 7325/7224  &            &         & & &\\\hline
kind2                &               &  851/737   &            &         & & &\\\hline
ldv-ant-med          &               &            & 10/10      &  79/79  & & &\\\hline
ldv-arrays           &               &            & 3/2        & 822/546 & & &\\\hline
llreve               &  44/44        &      43/42 & 31/31      &         & & &\\\hline
quic3                &               &            & 43/43      &         & & &\\\hline
ringen~(\textit{updated}) &          &            &            &         & & 454/440 &\\\hline
sally                &               &            &            &         & 177/174 & &\\\hline
seahorn              &  3379/2812    &     68/66  &            &         & &\\\hline
solidity~(\textit{new}) &            &            &            &         & & & 3571/3548\\\hline
sv-comp              &  3150/2930    &  1643/1169 & 79/73      & 856/780 & & &\\\hline
synth/nay-horn       &               &    119/114 &            &         & & &\\\hline
synth/semgus         &               &            &            & 5371/4839 & & &\\\hline
tricera              &  405/405      &  4/4       &            &         & & &\\\hline
tricera/adt-arrays~(\textit{new}) &  &            &            &         & & & 156/156\\\hline
ultimate~(\textit{new})  &           &  8/8       &            &  21/21  & & &\\\hline
vmt                  &  906/803      &            &            &         & 99/98 & &\\\hline\hline
{\bf total }       & 8421/{\bf 7501} & 10331/{\bf 9628} 
                   &   495/{\bf 488} & 7174/{\bf 6290} 
                   &   504/{\bf 498} &  521/{\bf  507} & 3727/{\bf 3704 }\\\hline
\end{tabular}
\caption{Summary of benchmarks (total/unique).}
\label{tab:bench_summary}
\end{table}

\clearpage
\subsection{Processing benchmarks with Algebraic Data Types and Reals}
For benchmarks containing either ADTs or Reals, no processing is done. All benchmarks submitted to the ADT-nonlin track were already processed using the \textsc{RInGen} tool~\cite{kostyukov2021finite} to encode all theory symbols using ADTs. The benchmarks submitted to the LRA-TS and LIA-nonlin-Arrays-nonrecADT track were already in compliance with the CHC-COMP format.

\subsection{Rating and Selection}
\label{subsub:selBench}
This section describes the procedure used to select benchmarks for the
competition.

For the LIA-lin-Arrays and LRA-TS tracks, consisting of a small amount
of benchmarks, all unique benchmarks were selected.

In all other tracks, consisting of (i) either a large amount of benchmarks
(that is, \LIAlin, \LIAnonlin, \LIAarrnonlin, and \ADTarrnonlin), or 
(ii) too few repositories (that is, \ADTnonlin, where we need to balance 
between the repositories), we followed a procedure similar to the
past editions of the competition aiming at selecting a representative subset
of the available benchmarks.
In particular, we estimated how ``easy" the benchmarks were and picked
a mix of ``easy" and ``hard" instances.
We say that a benchmark in a track is ``easy" if it is solved by both
the winner and the runner-up solvers in the corresponding track in 
CHC-COMP 2021, within a small time interval.
Each benchmark was rated A/B/C/D based on how difficult the previous
competition winners found them. A rating of ``A" is given if both solvers
solved the benchmark, ``B" if only the winner solved it, ``C" if only the
runner-up solved it, and ``D" if neither solved it, within the set timeout.
The timeout was selected based on the solver:
\begin{itemize}
    \item \textbf{Spacer} was run with a timeout of 5s for all configurations, and
    \item \textbf{Eldarica} was run with a timeout of 10s for all configurations. 
\end{itemize}
Eldarica was run with a higher timeout to compensate for the delay caused  by JVM start-up. All solvers were run using the same binaries and configurations submitted for CHC-COMP 2021. For the newly introduced LIA-nonlin-Arrays-nonrecADT track, the winners from ADT-nonlin track of CHC-COMP 2021 were used.

Once we labelled each benchmark from a repository $r$, we decided the maximum number of instances, $N_r$, to take from the repository. $N_r$ number was decided based on the total number of unique benchmarks and our knowledge about the benchmarks in repository $r$. We picked at most $0.2\cdot N_r$ benchmarks each with ratings A, B, and C. Then, we picked $0.4\cdot N_r$ benchmarks with rating D. If we did not find enough benchmarks with rating A, we picked the rest of the benchmarks equally from ratings B and C. If we did not find enough benchmarks with rating B or C, we pick the remaining benchmarks from rating D. This way, we obtained a mix of ``easy" and ``hard" benchmarks with a bias towards benchmarks that were not easily solved by either of the best solvers from the previous year's competition. The number of instances with each rating is given in~\cref{tab:bench_rating_1,tab:bench_rating_2}. The number of instances picked from each repository is given in Table~\ref{tab:inst_picked}. To pick \texttt{<num>} benchmarks of rating \texttt{<Y>}, we used the command

\begin{verbatim}
    > cat <rating-Y-benchmark-list> | sort -R | head -n <num>
\end{verbatim}

The final set of benchmarks selected for CHC-COMP 2022 can be found in the github repository
\url{https://github.com/chc-comp/chc-comp22-benchmarks}, and on StarExec in the public space \texttt{CHC/CHC-COMP/chc-comp22}.

\begin{table}[!ht]
\centering
\begin{tabular}{@{\hspace{2pt}}l                     @{\hspace{15pt}}
  r@{\hspace{3pt}}r@{\hspace{3pt}}r@{\hspace{3pt}}r @{\hspace{25pt}}
  r@{\hspace{3pt}}r@{\hspace{3pt}}r@{\hspace{3pt}}r @{\hspace{25pt}}
  r@{\hspace{3pt}}r@{\hspace{3pt}}r@{\hspace{3pt}}r}
\hline
&  \multicolumn{4}{c}{\LIAlin~~~~~}      
&  \multicolumn{4}{c}{\LIAnonlin~~~~~}  
&  \multicolumn{4}{c}{\LIAarrnonlin}\\
 Repository & \#A& \#B& \#C& \#D& \#A& \#B& \#C& \#D& \#A& \#B& \#C & \#D \\\hline\hline
aeval                & 11&  9&~~~~~2& 32& &&&& &&&\\\hline
aeval-unsafe         & 11&  5& 0& 38& &&&& &&&\\\hline
eldarica-misc        & 84& 39& 2& 11&  9& 26&~~~~~1& 30&&&\\\hline
extra-small-lia      & 13& 22& 3& 17& &&&& &&&\\\hline
hcai                 & 77&  5& 0&  5& 71& 41& 0& 19& 12& 6&~~~~~1& 6\\\hline
hopv                 & 47&  1& 0&  0& 46& 14& 7&  0&    &   &   &  \\\hline
jayhorn              & 73&  0& 0&  0& 1870& 3441& 1& 1912 & & & \\\hline
kind2                &    &    &   &   &   54&  660& 0&   23 & & & \\\hline
ldv-ant-med          &    &    &   &   &      &      &   &      & 0 & 15& 0& 64\\\hline
ldv-arrays           &    &    &   &   &      &      &   &      & 0 & 112& 0& 434\\\hline
llreve               & 34&  4& 2&  4&   10& 20& 1& 11&  & & &\\\hline
seahorn              & 678& 1306& 1& 827&  28& 25& 0& 13& & &\\\hline
sv-comp              & 2361& 475& 1&  93& 309& 766& 5& 89& 245& 254& 1& 280\\\hline
synth/nay-horn       &      &     &   &    &  25&  45& 0& 44&     &    &   &\\\hline
synth/semgus         &      &     &  &     &   &&& & 136& 2386& 0& 2317\\\hline
tricera/svcomp20     & 15& 27& 1& 362 &  4& 0& 0& 0& &&\\\hline
ultimate             &    &    &   &     &  0& 0& 0& 8& 0&0&0&21\\\hline
vmt                  & 26& 680& 0& 97 &    &   &   &  &   &  &  & \\\hline\hline
{\bf total }         &  3430& 2573& 12& 1486 
                     &  2426& 5038& 15& 2149  
                     &   393& 2773&  2& 3122\\\hline
\end{tabular}
\caption{The number of unique benchmarks with ratings A/B/C/D - Tracks: \LIAlin, \LIAnonlin, and \LIAarrnonlin.}
\label{tab:bench_rating_1}
\end{table}
\begin{table}[!ht]
\centering
\begin{tabular}{l                                   @{\hspace{20pt}}
  r@{\hspace{3pt}}r@{\hspace{3pt}}r@{\hspace{3pt}}r @{\hspace{25pt}}
  r@{\hspace{3pt}}r@{\hspace{3pt}}r@{\hspace{3pt}}r}
  \hline
Repository & \multicolumn{4}{l}{\ADTnonlin} &\multicolumn{4}{l}{LIA-nonlin-}\\
           & & & & &\multicolumn{4}{c}{Arrays-nonrecADT}\\\hline\hline
 Repository & \#A& \#B& \#C& \#D& \#A& \#B& \#C & \#D \\\hline\hline
adt-purified       &    5&   32&  1&  29 &      &      &    &    \\\hline
ringen             &   11&   17&  3& 409 &      &      &    &    \\\hline
solidity           &      &      &    &     & 1033& 1849& 68& 598\\\hline
tricera/adt-arrays &      &      &    &     &    2&   29&  0& 125\\\hline\hline
{\bf total }       &   16&   49&  4& 438 & 
                     1035& 1878& 68& 723 \\\hline
\end{tabular}
\caption{The number of unique benchmarks with ratings A/B/C/D -- Tracks: \ADTnonlin, and \ADTarrnonlin.}
\label{tab:bench_rating_2}
\end{table}

\begin{table}[!htbp]
\centering
\begin{tabular}{
@{\hspace{2pt}}l@{\hspace{0pt}}
@{\hspace{3pt}}r@{\hspace{3pt}}
@{\hspace{3pt}}r@{\hspace{3pt}}
@{\hspace{3pt}}r@{\hspace{3pt}}
@{\hspace{3pt}}r@{\hspace{3pt}}
@{\hspace{3pt}}r@{\hspace{3pt}}}
\hline
Repository 
&  \multirow{1}{1.4cm}{\LIAlin}     
&  \multirow{2}{1.4cm}{\LIAnonlin}   
&  \multirow{3}{1.4cm}{\LIAarrnonlin} 
&  \multirow{2}{1.4cm}{\ADTnonlin}
&  \multirow{4}{1.7cm}{\ADTarrnonlin}\\
                     &                 &              &           &         &\\
                     &                 &              &           &         &\\
                     &                 &              &           &         &\\\hline\hline
adt-purified         &                 &              &           &         67/52 & \\\hline
aeval                & 30/30 &                      &           &         &\\\hline
aeval-unsafe         & 30/30 &                      &           &         &\\\hline
eldarica-misc        & 45/31 & 30/30   &         &         &\\\hline
extra-small-lia      & 30/30 &        &         &         &\\\hline
hcai                 & 45/19 & 60/43  &  15/13  &         &\\\hline
hopv                 &  30/7 & 30/18   &         &         &\\\hline
jayhorn              &  30/6 & 90/90   &         &         &\\\hline
kind2                &       & 90/59   &         &         &\\\hline
ldv-ant-med          &       &         &  60/60  &         &\\\hline
ldv-arrays           &       &         &  90/90  &         &\\\hline
llreve               &  30/16& 45/30   &        &         &\\\hline
ringen               &                 &             &                     & 134/131 & \\\hline
seahorn              & 90/90 & 45/31   &            &         &\\\hline
solidity             &                  &            &            &         & 312/310 \\\hline
sv-comp              & 90/90 & 90/90   & 135/135 &  &\\\hline
synth/nay-horn       &       & 60/60      &            &         &\\\hline
synth/semgus         &       &           &  135/135 &         &\\\hline
tricera/svcomp20     & 60/60 &  3/0    &        &         & \\\hline
tricera/adt-arrays   &       &            &            &    & 156/155  \\\hline
ultimate             &       &  6/5          &  15/15   &         &\\\hline
vmt                  & 90/90 &        &            &         &\\\hline\hline
{\bf total }         & 600/{\bf 499 } &  549/{\bf 456 } & 450/{\bf 448 } & 201/{\bf 183 } & 468/{\bf 465 }\\\hline

\end{tabular}
\caption{The number of benchmarks to select and the number of selected benchmarks from each repository.}
\label{tab:inst_picked}
\end{table}

\clearpage
\section{Solvers}
\label{sec:solvers}

Six solvers were submitted to CHC-COMP 2022: five 
competing solvers, and one solver \textit{hors concours} 
(Spacer is co-developed by Hari Govind V K, who is
one of the Organizers of the CHC-COMP 2022).

Table~\ref{tab:solvers} lists the submitted solvers together with
the configurations used to run them on the competition tracks.
Detailed descriptions of the solvers are provided in
Section~\ref{sec:tools}. The binaries of the solvers are
available on StarExec in the public space 
\texttt{CHC/CHC-COMP/chc-comp22}. 

\medskip
\begin{table}[!h]
    \centering
    \begin{tabular}{p{1.3cm}
    p{1.3cm}p{1.3cm}p{1.3cm}p{1.3cm}
    p{1.4cm}p{1.48cm}p{1.3cm}p{1.9cm}}
    \hline
    \multirow{4}{*}{\textbf{Solver}} & 
    \LIAlin & \LIAnonlin & \LIAarrlin & \LIAarrnonlin &
    \LRATSlin & \LRATSpar &
    \ADTnonlin & \ADTarrnonlin\\\hline\hline
    Eldarica      &
                  \texttt{def} &
                  \texttt{def} &
                  \texttt{def} &
                  \texttt{def} & 
                  $\Box$       &
                  $\Box$       &
                  \texttt{def} &
                  \texttt{def}\\\hline
    Golem         &
                  \texttt{lia- lin}    &
                  \texttt{lia- nonlin} & 
                  $\Box$               &
                  $\Box$               &
                  \texttt{lra-ts}      &
                  \texttt{lra-ts}      & 
                  $\Box$               &
                  $\Box$               \\\hline
    RInGen        & 
                  $\Box$           &
                  $\Box$           &
                  $\Box$           &
                  $\Box$           &
                  $\Box$           &
                  $\Box$           &
                  \texttt{vampire} &
                  $\Box$           \\\hline
    Ultimate      
    TreeAutomizer &
                  \texttt{default} &
                  \texttt{default} &
                  \texttt{default} &
                  \texttt{default} &
                  \texttt{default} &
                  \texttt{default} &
                  \texttt{default} &
                  \texttt{default} \\\hline
    Ultimate      
    Unihorn       &
                  \texttt{default} &
                  \texttt{default} &
                  \texttt{default} &
                  \texttt{default} &
                  \texttt{default} &
                  \texttt{default} &
                  $\Box$           &
                  $\Box$           \\\hline
    \rowcolor{lightgray} Spacer        &
                  \texttt{LIA- LIN}            &
                  \texttt{LIA- NONLIN}         & 
                  \texttt{LIA- LIN- ARRAYS}    &
                  \texttt{LIA- NONLIN- ARRAYS} &
                  \texttt{LRA- TS}             &
                  \texttt{LRA- TS}             & 
                  \texttt{ADT- LIN}            &
                  \texttt{ADT- NONLIN}\\\hline
    \end{tabular}
    \caption{Submitted solvers and configurations 
    used in the competition track; `$\Box$' denotes
    that the solver did not enter the competition 
    in that track.
    The configuration names have been taken as is 
    from solver submissions.}
    \label{tab:solvers}
\end{table}
 

\section{Results}
\label{sec:results}

The results of the CHC-COMP 2022 are presented in 
Table~\ref{tab:results}. 

Eldarica and Ultimate TreeAutomizer were the only solvers 
that entered the competition in the `\ADTarrnonlin' track.
Ultimate TreeAutomizer did not provide any \textit{sat} and
\textit{unsat} result, therefore it is not included as 2nd
classified in the the final ranking.

Detailed results for the eight tracks are provided in 
Appendix~\ref{app:detres} 

\begin{table}[!h]
    \centering
    \begin{tabular}{l
    p{1.29cm}p{1.29cm}p{1.29cm}p{1.29cm}
    p{1.35cm}p{1.48cm}p{1.25cm}p{1.9cm}}
    \hline
    & 
    \LIAlin & \LIAnonlin & \LIAarrlin & \LIAarrnonlin &
    \LRATSlin & \LRATSpar &
    \ADTnonlin & \ADTarrnonlin\\\hline\hline
    \textbf{Winner} &
        \textbf{Golem}    &
        \textbf{Golem}    &
        \textbf{Eldarica} &
        \textbf{Eldarica} & 
        \textbf{Golem}    &
        \textbf{Golem}    &
        \textbf{RInGen}   &
        \textbf{Eldarica} \\\hline
    2nd place &
        Eldarica               &
        Eldarica               & 
        Ultimate Unihorn       &
        Ultimate Unihorn       &
        Ultimate TreeAutomizer &
        Ultimate TreeAutomizer &
        Eldarica               & 
        \\\hline
    3rd place & 
        Ultimate Unihorn       &
        Ultimate Unihorn       &
        Ultimate TreeAutomizer &
        Ultimate TreeAutomizer &
        Ultimate Unihorn       &
        Ultimate Unihorn       &
        &
        \\\hline
    \end{tabular}
    \caption{Results of the competition.}
    \label{tab:results}
\end{table}

\subsection{Observed Issues and Fixes during the 
Competition Runs}

In the competition runs of the \LIAarrlin and \LRATSlin
tracks we detected 4 and 17 inconsistent results,
respectively.
\textbf{Ultimate Unihorn} reported \textit{sat}, while
other competing tools reported \textit{unsat}.
In particular, in the \LIAarrlin track, we observed that
both Eldarica and Ultimate TreeAutomizer reported
\textit{unsat} on 2 out of 4 \textit{sat} results 
reported by Ultimate Unihorn, while on the other 2 
\textit{sat} results Ultimate TreeAutomizer reported 
\textit{unsat} and Eldarica \textit{unknown}.
In the \LRATSlin track, Golem reported 17 \textit{unsat}
results, while Ultimate TreeAutomizer reported 12 \textit{unsat} 
and 5 \textit{unknown} results, respectively. 

The inconsistencies were detected on March 24, 2022.
We informed the authors of Ultimate Unihorn on March 25,
2022 by sending them six benchmarks on which we detected
the inconsistencies: two \LIAarrlin benchmarks and four
\LRATSlin benchmarks.
The authors submitted a fixed version of their tool on
March 28, 2022.

The results presented in this report were produces using
the fixed version. 
In Table~\ref{tab:unihorn-inconsistencies} we report
the results before and after the fixes. 

\begin{table}[!h]
    \centering
    \begin{tabular}{p{2cm} p{1cm}p{1.5cm} p{1cm}p{1cm} }
    \hline
    Ultimate  & \multicolumn{2}{c}{\LIAarrlin} &
                \multicolumn{2}{c}{\LRATSlin}  \\
    Unihorn   & \#\textit{sat} & \#\textit{unsat} &
                \#\textit{sat} & \#\textit{unsat} \\\hline\hline
    original  & 283 & 66 & 134 & 22 \\\hline
    fixed     & 137 & 67 &  65 & 38 \\\hline
    \end{tabular}
    \caption{Results produced by Ultimate Unihorn 
    before (original) and after the bug fixes (fixed).}
    \label{tab:unihorn-inconsistencies}
\end{table}

\clearpage
\section{Conclusions and Final Remarks}
\label{sec:conclusions}

We would like to congratulate the winners of the CHC-COMP 2022
(in alphabetical order): 
\textbf{Eldarica} (winner of the \LIAarrlin and \LIAarrlin 
tracks, and the newly introduced \ADTarrnonlin track),
\textbf{Golem} (winner of the \LIAlin, \LIAnonlin, \LRATSlin,
and \LRATSpar tracks), and \textbf{RInGen} (winner of the 
\ADTnonlin track).

We conclude with a few remarks on the open issues that should be 
discussed and addressed in the future editions of the CHC-COMP,
which are based on our experience with running the competition
and the observations made by the HCVS 2022 participants in the 
follow-up discussion we had after the presentation of the
competition results.

\begin{itemize}
\item \textbf{Validation of results}.
The ability of solvers to generate models and counterexamples is
a recurrent request by our community members (this issue has been
already discussed in the previous editions, see~\cite{chccomp21}).
Thus, to encourage the developers to introduce this feature 
(some solvers already provide it), we could begin, 
as proposed in the CHC-COMP 2021 report~\cite{chccomp21},
by introducing new tracks where this feature is taken into
consideration in the computation of the score. For instance,
the score could be weighted according to type of witness 
provided by the solver to support its result.

\item \textbf{Status of benchmarks}.
In order to assess the correctness of the result provided by the
solvers, each submitted benchmark should explicitly declare the
expected result of the satisfiability problem.
We propose to use the \texttt{( set-info}
$\langle keyword\rangle$ $\langle \textit{attr-value} \rangle$ 
\texttt{)} command with the \texttt{:status} as \textit{keyword},
and either \texttt{sat} or \texttt{unsat} as \textit{attr-value}.

\item \textbf{The \LRATSlin and \LRATSpar tracks}.
As already discussed in~\cite{chccomp21}, also in this edition, 
no solver requiring the syntactic restriction on the form of
the clauses included in the \LRATSlin track has been submitted.
Hence, we propose to discontinue the \LRATSlin and \LRATSpar
tracks starting from the CHC-COMP 2023, and to add more general
LRA tracks, such as LRA-lin and LRA-nonlin.

\item \textbf{The \ADTnonlin and \ADTarrnonlin tracks}.
As already discussed in~\cite{chccomp21}, the syntactic restrictions
on the form of the clauses in \ADTnonlin track were meant to attract
more solvers to enter the competition. 
Nowadays, there is an increasing interest in developing techniques 
for solving CHCs with constraints over ADTs and LIA/LRA~\cite{kostyukov2021finite,DBLP:journals/logcom/AngelisFPP22,DBLP:journals/pacmpl/KSG22}, thereby 
increasing the availability of tools supporting these theories.
In this regard, CHC-COMP 2022 introduced the \ADTarrnonlin track,
which combines nonrecursive ADTs with LIA and Arrays, but only 
two solvers, that is, Eldarica and Spacer, were able to enter the
competition in this track.
We propose to try a more gradual combination of such theories,
such as ADTs with LIA and ADTs with Arrays.

\item \textbf{Generation of the benchmark suite for the 
competition runs}.
The process for constructing the benchmark suite is based
on the evaluation of the hardness of the encoded problems.
This evaluation is performed by running the winners of the 
previous edition for a very limited amount of time (see
Section~\ref{subsub:selBench})
In the follow-up discussion at HCVS, it has been suggested
to increase the amount of time given to the solvers to get
a more accurate evaluation of the hardness of the benchmarks.

\end{itemize}

Finally, we would also to stress the fact that \textbf{a bigger set
of benchmarks are needed}. Besides submitting their tools, all
participants are invited to contribute with new benchmarks.

\section{Solver Descriptions}
\label{sec:tools}

The tool descriptions in this section were contributed by the
participants, and the copyright on the texts remains with the 
individual authors.

\newcommand{\toolname}[1]{\subsection{#1}}
\newcommand{\toolsubmitter}[2]{\noindent #1\\#2\par\smallskip}
\newcommand{\toolalgorithm}{\paragraph{Algorithm.}}
\newcommand{\toolarchitecture}{\paragraph{Architecture and Implementation.}}
\newcommand{\toolconfiguration}{\paragraph{Configuration in CHC-COMP 2022.}}
\newcommand{\toolnew}{\paragraph{New Features in CHC-COMP 2022.}}
\newcommand{\toollink}[2]{\par\bigskip\noindent\url{#1}\\#2}

\toolname{Eldarica v2.0.8}


\toolsubmitter{Hossein Hojjat}{University of Tehran, Iran}
\toolsubmitter{Philipp R\"ummer}{University of Regensburg, Germany}

\toolalgorithm

Eldarica~\cite{FMCAD2018HojjatRummer} is a Horn solver applying
classical algorithms from model checking: predicate abstraction and
counterexample-guided abstraction refinement (CEGAR).  Eldarica can
solve Horn clauses over linear integer arithmetic, arrays, algebraic
data-types, bit-vectors, and the theory of heaps.  It can process Horn
clauses and programs in a variety of formats, implements sophisticated
algorithms to solve tricky systems of clauses without diverging, and
offers an elegant API for programmatic use.

\toolarchitecture

Eldarica is entirely implemented in Scala, and only depends on Java or
Scala libraries, which implies that Eldarica can be used on any
platform with a JVM. For computing abstractions of systems of Horn
clauses and inferring new predicates, Eldarica invokes the SMT solver
Princess~\cite{princess08} as a library.

\toolconfiguration

Eldarica is in the competition run with the option \verb!-portfolio!,
which enables a simple portfolio mode: three instances of the solver
are run in parallel,
one with options \verb!-splitClauses:0 -abstract:off!,
one with options \verb!-splitClauses:1 -abstract:off!, and
one with default options.

\toollink{https://github.com/uuverifiers/eldarica}{BSD licence}

\toolname{\golem}


\toolsubmitter{Martin Blicha}{Universit\`{a} della Svizzera italiana, Switzerland}

\toolalgorithm
\golem{} is a CHC solver under active development that provides several backend engines implementing various interpolation-based model-checking algorithms.
It supports the theory of Linear Real or Integer Arithmetic and it is able to provide witnesses for both satisfiable and unsatisfiable CHC systems.
The three engines of \golem{} are:
\begin{itemize}
\item \texttt{lawi} is our re-implementation of the \textsc{Impact} algorithm~\cite{McMillan_2006}
\item \texttt{spacer} is our re-implementation of the \textsc{Spacer} algorithm~\cite{Komuravelli2016} and allows \golem{} to solve non-linear systems.
\item \texttt{tpa} is our new model-checking algorithm based on doubling abstractions using Craig interpolants~\cite{Blicha_2022}.
\end{itemize}

\toolarchitecture

\golem{} is implemented in C++ and built on top of the interpolating SMT solver \textsc{OpenSMT}~\cite{OpenSMT2} which is used for both satisfiability solving and interpolation. The only dependencies are those inherited from  \textsc{OpenSMT}: Flex, Bison and GMP libraries.

\toolnew
Compared to the previous year, \golem{} has two new backend engines, \texttt{spacer} and \texttt{tpa}.
Additionally, \golem{} now uses basic preprocessing to simplify the input clauses before handing them over to the backend engine.

\toolconfiguration
For LIA-nonlin track we used only \texttt{spacer} engine; the other engines cannot handle nonlinear system yet.

\texttt{\$ golem --logic QF\_LIA --engine spacer}

For LIA-lin and LRA-TS tracks, we used a trivial portfolio of all three engines running independently.

\toollink{https://github.com/usi-verification-and-security/golem}{MIT LICENSE}

\newcommand{\ourtool}{RInGen}
\newcommand{\cvc}{\textsc{CVC4}}
\newcommand{\vampire}{\textsc{Vampire}}

\toolname{\ourtool{} v1.2}


\toolsubmitter{Yurii Kostyukov}{JetBrains Research, Russia}
\toolsubmitter{Dmitry Mordvinov}{JetBrains Research, Russia}


\toolalgorithm

\ourtool{} is a \emph{R}egular \emph{In}variant \emph{Gen}erator and a first-order logic formula transformer.
\ourtool{} is based on the preprocessing approach presented
in~\cite{kostyukov2021finite}.
A system of constraint Horn clauses (CHCs) over algebraic datatypes (ADTs) is rewritten into a formula over uninterpreted function symbols. Crucial step is elimination of all disequalities, testers, and selectors from the clause bodies by introducing their Horn axioms. Then the satisfiability modulo theory of ADTs is reduced to satisfiability modulo theory of uninterpreted functions with equality (EUF) by replacing all ADT sorts with free sorts and and all constructors with free functions. After that, an off-the-shelf logical solver for many-sorted logic with quantifiers is called. As the proposed transformation gives a formula which is satisfiable modulo EUF iff the original CHC system is satisfiable modulo ADTs, the result of the logical solver is returned as is.

\toolarchitecture

\ourtool{} accepts input in the SMTLIB2 format and produces Horn clauses over pure ADT sorts in SMTLIB2 and Prolog.
It takes conditions with a property and checks if the property holds, returning SAT and the safe inductive invariant if it does or terminates with UNSAT if it does not.
We run \vampire{} as a backend many-sorted EUF solver.
\vampire{}~\cite{vampire} searches for both refutations and saturations of it's input problem, which gives us both SAT and UNSAT results for CHC systems.

\toolnew

The main feature of this year contribution is using our fine-tuned fork of \vampire{} as a backend.

\toolconfiguration

The tool is run with the following arguments:\\
\centerline{\texttt{--timelimit \$tl -q -o "\$2/" solve -s vampire --path "\$input" -t --no-isolation}.}
The tool runs our fork\footnote{\url{https://github.com/Columpio/Vampire/tree/chc-comp22}} of \vampire{} (using \texttt{vampire -{}-mode chccomp}) tuned for CHC problems as a backend solver.

\toollink{https://github.com/Columpio/RInGen/releases/tag/chccomp22}{BSD 3-Clause License}



\toolname{Ultimate TreeAutomizer 0.2.2-dev-f165340}


\toolsubmitter{Matthias Heizmann}{University of Freiburg, Germany}
\toolsubmitter{Daniel Dietsch}{University of Freiburg, Germany}
\toolsubmitter{Jochen Hoenicke}{University of Freiburg, Germany}
\toolsubmitter{Alexander Nutz}{University of Freiburg, Germany}
\toolsubmitter{Andreas Podelski}{University of Freiburg, Germany}
\toolsubmitter{Frank Sch\"ussele}{University of Freiburg, Germany}

\toolalgorithm

The \textsc{Ultimate TreeAutomizer} solver implements an approach that is based on tree automata~\cite{journals/corr/abs-1907-03998}.
In this approach potential counterexamples to satisfiability are considered as a regular set of trees.
In an iterative \nobreak{CEGAR} loop we analyze potential counterexamples.
Real counterexamples lead to an \textit{unsat} result.
Spurious counterexamples are generalized to a regular set of spurious counterexamples
and subtracted from the set of potential counterexamples that have to be considered.
In case we detected that all potential counterexamples are spurious, the result is \textit{sat}.
The generalization above is based on tree interpolation and
regular sets of trees are represented as tree automata.


\toolarchitecture

\textsc{TreeAutomizer} is a toolchain in the 
\textsc{Ultimate} framework.
This toolchain first parses the CHC input and then runs the \texttt{treeautomizer} plugin which
implements the above mentioned algorithm.
We obtain tree interpolants from the SMT solver SMTInterpol%
\footnote{\url{https://ultimate.informatik.uni-freiburg.de/smtinterpol/}}%
~\cite{cade/HoenickeS18}.
For checking satisfiability, we use the
and Z3 SMT solver%
\footnote{\url{https://github.com/Z3Prover/z3}}%
.
The tree automata are implemented in \textsc{Ultimate}'s automata library%
\footnote{\url{https://ultimate.informatik.uni-freiburg.de/automata_library}}%
.
The \textsc{Ultimate} framework is written in Java and build upon the Eclipse Rich Client Platform (RCP). The source code is available at
GitHub\footnote{\url{https://github.com/ultimate-pa/}}.


\toolconfiguration

Our StarExec archive for the competition is shipped with the \texttt{bin/starexec\_run\_default}
shell script calls the \textsc{Ultimate} command line interface with the
\texttt{TreeAutomizer.xml} toolchain file and
the \texttt{TreeAutomizerHopcroftMinimization.epf} settings file.
Both files can be found in toolchain (resp. settings) folder of \textsc{Ultimate}'s repository.


\toollink{https://ultimate.informatik.uni-freiburg.de/}{LGPLv3 with a linking exception for Eclipse RCP}


\toolname{Ultimate Unihorn 0.2.2-dev-f165340}


\toolsubmitter{Matthias Heizmann}{University of Freiburg, Germany}
\toolsubmitter{Daniel Dietsch}{University of Freiburg, Germany}
\toolsubmitter{Jochen Hoenicke}{University of Freiburg, Germany}
\toolsubmitter{Alexander Nutz}{University of Freiburg, Germany}
\toolsubmitter{Andreas Podelski}{University of Freiburg, Germany}
\toolsubmitter{Frank Sch\"ussele}{University of Freiburg, Germany}

\toolalgorithm

\textsc{Ultimate Unihorn} reduces the satisfiability problem for a set of constraint Horn clauses
to a software verfication problem.
In a first step \textsc{Unihorn} applies a 
yet unpublished translation in which the constraint Horn clauses
are translated into a recursive program
that is nondeterministic and
whose correctness is specified by an assert statement
The program is correct (i.e., no execution violates the assert statement)
if and only if the set of CHCs is satisfiable.
For checking whether the recursive program satisfies its specification,
Unihorn uses \textsc{Ultimate Automizer}~\cite{tacas/HeizmannCDGHLNM18}
which implements an automata-based approach to software verification~\cite{cav/HeizmannHP13}.


\toolarchitecture

\textsc{Ultimate Unihorn} is a toolchain in the 
\textsc{Ultimate} framework.
This toolchain first parses the CHC input and then runs the \texttt{chctoboogie} plugin which
does the translation from CHCs into a recursive program.
We use the Boogie
language to represent that program.
Afterwards the default toolchain for verifying a recursive Boogie programs by \textsc{Ultimate Automizer} is applied.
The \textsc{Ultimate} framework shares the libraries for handling SMT formulas with the SMTInterpol SMT solver.
While verifying a program, \textsc{Ultimate Automizer} needs SMT solvers
for checking satisfiability,
for computing Craig interpolants and
for computing unsatisfiable cores.
The version of \textsc{Unihorn} that participated in the competition
used the SMT solvers SMTInterpol%
\footnote{\url{https://ultimate.informatik.uni-freiburg.de/smtinterpol/}}%
and Z3%
\footnote{\url{https://github.com/Z3Prover/z3}}%
.
The \textsc{Ultimate} framework is written in Java and build upon the Eclipse Rich Client Platform (RCP). The source code is available at
GitHub\footnote{\url{https://github.com/ultimate-pa/}}.


\toolconfiguration

Our StarExec archive for the competition is shipped with the \texttt{bin/starexec\_run\_default}
shell script calls the \textsc{Ultimate} command line interface with the
\texttt{AutomizerCHC.xml} toolchain file and
the \texttt{chccomp-Unihorn\_Default.epf} settings file.
Both files can be found in toolchain (resp. settings) folder of \textsc{Ultimate}'s repository.


\toollink{https://ultimate.informatik.uni-freiburg.de/}{LGPLv3 with a linking exception for Eclipse RCP}

\bibliographystyle{eptcs}
\bibliography{biblio}

\begin{thebibliography}{10}
\providecommand{\bibitemdeclare}[2]{}
\providecommand{\surnamestart}{}
\providecommand{\surnameend}{}
\providecommand{\urlprefix}{Available at }
\providecommand{\url}[1]{\texttt{#1}}
\providecommand{\href}[2]{\texttt{#2}}
\providecommand{\urlalt}[2]{\href{#1}{#2}}
\providecommand{\doi}[1]{doi:\urlalt{http://dx.doi.org/#1}{#1}}
\providecommand{\bibinfo}[2]{#2}

\bibitemdeclare{misc}{BarFT-SMTLIB}
\bibitem{BarFT-SMTLIB}
\bibinfo{author}{Clark \surnamestart Barrett\surnameend},
  \bibinfo{author}{Pascal \surnamestart Fontaine\surnameend} \&
  \bibinfo{author}{Cesare \surnamestart Tinelli\surnameend}
  (\bibinfo{year}{2016}): \emph{\bibinfo{title}{{The Satisfiability Modulo
  Theories Library (SMT-LIB)}}}.
\newblock \bibinfo{howpublished}{{\tt www.SMT-LIB.org}}.

\bibitemdeclare{incollection}{BGMR2015}
\bibitem{BGMR2015}
\bibinfo{author}{Nikolaj \surnamestart Bj{\o}rner\surnameend},
  \bibinfo{author}{Arie \surnamestart Gurfinkel\surnameend},
  \bibinfo{author}{Ken \surnamestart McMillan\surnameend} \&
  \bibinfo{author}{Andrey \surnamestart Rybalchenko\surnameend}
  (\bibinfo{year}{2015}): \emph{\bibinfo{title}{Horn Clause Solvers for Program
  Verification}}.
\newblock In \bibinfo{editor}{Lev~D. \surnamestart Beklemishev\surnameend},
  \bibinfo{editor}{Andreas \surnamestart Blass\surnameend},
  \bibinfo{editor}{Nachum \surnamestart Dershowitz\surnameend},
  \bibinfo{editor}{Bernd \surnamestart Finkbeiner\surnameend} \&
  \bibinfo{editor}{Wolfram \surnamestart Schulte\surnameend}, editors: {\sl
  \bibinfo{booktitle}{Fields of Logic and Computation II: Essays Dedicated to
  Yuri Gurevich on the Occasion of His 75th Birthday}},
  \bibinfo{publisher}{Springer International Publishing},
  \bibinfo{address}{Cham}, pp. \bibinfo{pages}{24--51},
  \doi{10.1007/978-3-319-23534-9\_2}.

\bibitemdeclare{inproceedings}{Blicha_2022}
\bibitem{Blicha_2022}
\bibinfo{author}{Martin \surnamestart Blicha\surnameend},
  \bibinfo{author}{Grigory \surnamestart Fedyukovich\surnameend},
  \bibinfo{author}{Antti E.~J. \surnamestart Hyv{\"a}rinen\surnameend} \&
  \bibinfo{author}{Natasha \surnamestart Sharygina\surnameend}
  (\bibinfo{year}{2022}): \emph{\bibinfo{title}{Transition Power Abstractions
  for Deep Counterexample Detection}}.
\newblock In \bibinfo{editor}{Dana \surnamestart Fisman\surnameend} \&
  \bibinfo{editor}{Grigore \surnamestart Rosu\surnameend}, editors: {\sl
  \bibinfo{booktitle}{Tools and Algorithms for the Construction and Analysis of
  Systems}}, \bibinfo{publisher}{Springer International Publishing},
  \bibinfo{address}{Cham}, pp. \bibinfo{pages}{524--542},
  \doi{10.1007/978-3-030-99524-9\_29}.

\bibitemdeclare{article}{DFGHPP2021}
\bibitem{DFGHPP2021}
\bibinfo{author}{Emanuele \surnamestart De~Angelis\surnameend},
  \bibinfo{author}{Fabio \surnamestart Fioravanti\surnameend},
  \bibinfo{author}{John~P. \surnamestart Gallagher\surnameend},
  \bibinfo{author}{Manuel~V. \surnamestart Hermenegildo\surnameend},
  \bibinfo{author}{Alberto \surnamestart Pettorossi\surnameend} \&
  \bibinfo{author}{Maurizio \surnamestart Proeitti\surnameend}
  (\bibinfo{year}{2021}): \emph{\bibinfo{title}{Analysis and Transformation of
  Constrained Horn Clauses for Program Verification}}.
\newblock {\sl \bibinfo{journal}{Theory and Practice of Logic Programming}}, p.
  \bibinfo{pages}{1--69}, \doi{10.1017/S1471068421000211}.

\bibitemdeclare{article}{DBLP:journals/logcom/AngelisFPP22}
\bibitem{DBLP:journals/logcom/AngelisFPP22}
\bibinfo{author}{Emanuele \surnamestart {De Angelis}\surnameend},
  \bibinfo{author}{Fabio \surnamestart Fioravanti\surnameend},
  \bibinfo{author}{Alberto \surnamestart Pettorossi\surnameend} \&
  \bibinfo{author}{Maurizio \surnamestart Proietti\surnameend}
  (\bibinfo{year}{2022}): \emph{\bibinfo{title}{Satisfiability of constrained
  Horn clauses on algebraic data types: {A} transformation-based approach}}.
\newblock {\sl \bibinfo{journal}{J. Log. Comput.}}
  \bibinfo{volume}{32}(\bibinfo{number}{2}), pp. \bibinfo{pages}{402--442},
  \doi{10.1093/logcom/exab090}.

\bibitemdeclare{inproceedings}{journals/corr/abs-1907-03998}
\bibitem{journals/corr/abs-1907-03998}
\bibinfo{author}{Daniel \surnamestart Dietsch\surnameend},
  \bibinfo{author}{Matthias \surnamestart Heizmann\surnameend},
  \bibinfo{author}{Jochen \surnamestart Hoenicke\surnameend},
  \bibinfo{author}{Alexander \surnamestart Nutz\surnameend} \&
  \bibinfo{author}{Andreas \surnamestart Podelski\surnameend}
  (\bibinfo{year}{2019}): \emph{\bibinfo{title}{Ultimate TreeAutomizer
  {(CHC-COMP} Tool Description)}}.
\newblock In: {\sl \bibinfo{booktitle}{HCVS/PERR@ETAPS}}, {\sl
  \bibinfo{series}{{EPTCS}}} \bibinfo{volume}{296}, pp.
  \bibinfo{pages}{42--47}, \doi{10.4204/eptcs.296.7}.

\bibitemdeclare{inproceedings}{chccomp21}
\bibitem{chccomp21}
\bibinfo{author}{Grigory \surnamestart Fedyukovich\surnameend} \&
  \bibinfo{author}{Philipp \surnamestart R{\"{u}}mmer\surnameend}
  (\bibinfo{year}{2021}): \emph{\bibinfo{title}{Competition Report:
  {CHC-COMP-21}}}.
\newblock In \bibinfo{editor}{Hossein \surnamestart Hojjat\surnameend} \&
  \bibinfo{editor}{Bishoksan \surnamestart Kafle\surnameend}, editors: {\sl
  \bibinfo{booktitle}{Proceedings 8th Workshop on Horn Clauses for Verification
  and Synthesis, HCVS@ETAPS 2021, Virtual, 28th March 2021}}, {\sl
  \bibinfo{series}{{EPTCS}}} \bibinfo{volume}{344}, \bibinfo{publisher}{Open
  Publishing Association}, pp. \bibinfo{pages}{91--108},
  \doi{10.4204/EPTCS.344.7}.

\bibitemdeclare{inproceedings}{Gurfinkel2022}
\bibitem{Gurfinkel2022}
\bibinfo{author}{Arie \surnamestart Gurfinkel\surnameend}
  (\bibinfo{year}{2022}): \emph{\bibinfo{title}{Program Verification
  with Constrained Horn Clauses (Invited Paper)}}.
\newblock In \bibinfo{editor}{Sharon \surnamestart Shoham\surnameend} \&
  \bibinfo{editor}{Yakir \surnamestart Vizel\surnameend}, editors: {\sl
  \bibinfo{booktitle}{Computer Aided Verification}},
  \bibinfo{publisher}{Springer International Publishing},
  \bibinfo{address}{Cham}, pp. \bibinfo{pages}{19--29},
  \doi{10.1007/978-3-031-13185-1\_2}.

\bibitemdeclare{inproceedings}{tacas/HeizmannCDGHLNM18}
\bibitem{tacas/HeizmannCDGHLNM18}
\bibinfo{author}{Matthias \surnamestart Heizmann\surnameend},
  \bibinfo{author}{Yu{-}Fang \surnamestart Chen\surnameend},
  \bibinfo{author}{Daniel \surnamestart Dietsch\surnameend},
  \bibinfo{author}{Marius \surnamestart Greitschus\surnameend},
  \bibinfo{author}{Jochen \surnamestart Hoenicke\surnameend},
  \bibinfo{author}{Yong \surnamestart Li\surnameend},
  \bibinfo{author}{Alexander \surnamestart Nutz\surnameend},
  \bibinfo{author}{Betim \surnamestart Musa\surnameend},
  \bibinfo{author}{Christian \surnamestart Schilling\surnameend},
  \bibinfo{author}{Tanja \surnamestart Schindler\surnameend} \&
  \bibinfo{author}{Andreas \surnamestart Podelski\surnameend}
  (\bibinfo{year}{2018}): \emph{\bibinfo{title}{Ultimate Automizer and the
  Search for Perfect Interpolants - (Competition Contribution)}}.
\newblock In: {\sl \bibinfo{booktitle}{{TACAS} {(2)}}}, {\sl
  \bibinfo{series}{Lecture Notes in Computer Science}} \bibinfo{volume}{10806},
  \bibinfo{publisher}{Springer}, pp. \bibinfo{pages}{447--451},
  \doi{10.1007/978-3-319-89963-3\_30}.

\bibitemdeclare{inproceedings}{cav/HeizmannHP13}
\bibitem{cav/HeizmannHP13}
\bibinfo{author}{Matthias \surnamestart Heizmann\surnameend},
  \bibinfo{author}{Jochen \surnamestart Hoenicke\surnameend} \&
  \bibinfo{author}{Andreas \surnamestart Podelski\surnameend}
  (\bibinfo{year}{2013}): \emph{\bibinfo{title}{Software Model Checking for
  People Who Love Automata}}.
\newblock In: {\sl \bibinfo{booktitle}{{CAV}}}, {\sl \bibinfo{series}{Lecture
  Notes in Computer Science}} \bibinfo{volume}{8044},
  \bibinfo{publisher}{Springer}, pp. \bibinfo{pages}{36--52},
  \doi{10.1007/978-3-642-39799-8\_2}.

\bibitemdeclare{inproceedings}{cade/HoenickeS18}
\bibitem{cade/HoenickeS18}
\bibinfo{author}{Jochen \surnamestart Hoenicke\surnameend} \&
  \bibinfo{author}{Tanja \surnamestart Schindler\surnameend}
  (\bibinfo{year}{2018}): \emph{\bibinfo{title}{Efficient Interpolation for the
  Theory of Arrays}}.
\newblock In: {\sl \bibinfo{booktitle}{{IJCAR}}}, {\sl \bibinfo{series}{Lecture
  Notes in Computer Science}} \bibinfo{volume}{10900},
  \bibinfo{publisher}{Springer}, pp. \bibinfo{pages}{549--565},
  \doi{10.1007/978-3-319-94205-6\_36}.

\bibitemdeclare{inproceedings}{FMCAD2018HojjatRummer}
\bibitem{FMCAD2018HojjatRummer}
\bibinfo{author}{Hossein \surnamestart Hojjat\surnameend} \&
  \bibinfo{author}{Philipp \surnamestart R{\"{u}}mmer\surnameend}
  (\bibinfo{year}{2018}): \emph{\bibinfo{title}{The {ELDARICA} Horn Solver}}.
\newblock In: {\sl \bibinfo{booktitle}{2018 Formal Methods in Computer Aided
  Design, {FMCAD}}}, pp. \bibinfo{pages}{1--7},
  \doi{10.23919/FMCAD.2018.8603013}.

\bibitemdeclare{inproceedings}{OpenSMT2}
\bibitem{OpenSMT2}
\bibinfo{author}{Antti E.~J. \surnamestart Hyv{\"a}rinen\surnameend},
  \bibinfo{author}{Matteo \surnamestart Marescotti\surnameend},
  \bibinfo{author}{Leonardo \surnamestart Alt\surnameend} \&
  \bibinfo{author}{Natasha \surnamestart Sharygina\surnameend}
  (\bibinfo{year}{2016}): \emph{\bibinfo{title}{{OpenSMT2}: An {SMT} Solver for
  Multi-core and Cloud Computing}}.
\newblock In \bibinfo{editor}{Nadia \surnamestart Creignou\surnameend} \&
  \bibinfo{editor}{Daniel \surnamestart Le~Berre\surnameend}, editors: {\sl
  \bibinfo{booktitle}{Theory and Applications of Satisfiability Testing -- SAT
  2016}}, \bibinfo{publisher}{Springer International Publishing},
  \bibinfo{address}{Cham}, pp. \bibinfo{pages}{547--553},
  \doi{10.1007/978-3-319-40970-2\_35}.

\bibitemdeclare{article}{DBLP:journals/pacmpl/KSG22}
\bibitem{DBLP:journals/pacmpl/KSG22}
\bibinfo{author}{Hari Govind~V. \surnamestart K.\surnameend},
  \bibinfo{author}{Sharon \surnamestart Shoham\surnameend} \&
  \bibinfo{author}{Arie \surnamestart Gurfinkel\surnameend}
  (\bibinfo{year}{2022}): \emph{\bibinfo{title}{Solving constrained Horn
  clauses modulo algebraic data types and recursive functions}}.
\newblock {\sl \bibinfo{journal}{Proc. {ACM} Program. Lang.}}
  \bibinfo{volume}{6}(\bibinfo{number}{{POPL}}), pp. \bibinfo{pages}{1--29},
  \doi{10.1145/3498722}.

\bibitemdeclare{article}{Komuravelli2016}
\bibitem{Komuravelli2016}
\bibinfo{author}{Anvesh \surnamestart Komuravelli\surnameend},
  \bibinfo{author}{Arie \surnamestart Gurfinkel\surnameend} \&
  \bibinfo{author}{Sagar \surnamestart Chaki\surnameend}
  (\bibinfo{year}{2016}): \emph{\bibinfo{title}{{SMT}-based Model Checking For
  Recursive Programs}}.
\newblock {\sl \bibinfo{journal}{Formal Methods in System Design}}
  \bibinfo{volume}{48}(\bibinfo{number}{3}), pp. \bibinfo{pages}{175--205},
  \doi{10.1007/s10703-016-0249-4}.

\bibitemdeclare{inproceedings}{kostyukov2021finite}
\bibitem{kostyukov2021finite}
\bibinfo{author}{Yurii \surnamestart Kostyukov\surnameend},
  \bibinfo{author}{Dmitry \surnamestart Mordvinov\surnameend} \&
  \bibinfo{author}{Grigory \surnamestart Fedyukovich\surnameend}
  (\bibinfo{year}{2021}): \emph{\bibinfo{title}{Beyond the Elementary
  Representations of Program Invariants over Algebraic Data Types}}.
\newblock In: {\sl \bibinfo{booktitle}{Proceedings of the 42nd ACM SIGPLAN
  International Conference on Programming Language Design and Implementation}},
  \bibinfo{series}{PLDI 2021}, \bibinfo{publisher}{{ACM}}, p.
  \bibinfo{pages}{451--465}, \doi{10.1145/3453483.3454055}.

\bibitemdeclare{inproceedings}{McMillan_2006}
\bibitem{McMillan_2006}
\bibinfo{author}{Kenneth~L. \surnamestart McMillan\surnameend}
  (\bibinfo{year}{2006}): \emph{\bibinfo{title}{Lazy Abstraction with
  Interpolants}}.
\newblock In \bibinfo{editor}{Thomas \surnamestart Ball\surnameend} \&
  \bibinfo{editor}{Robert~B. \surnamestart Jones\surnameend}, editors: {\sl
  \bibinfo{booktitle}{Computer Aided Verification}},
  \bibinfo{publisher}{Springer Berlin Heidelberg}, \bibinfo{address}{Berlin,
  Heidelberg}, pp. \bibinfo{pages}{123--136}, \doi{10.1007/11817963\_14}.

\bibitemdeclare{article}{vampire}
\bibitem{vampire}
\bibinfo{author}{Alexandre \surnamestart Riazanov\surnameend} \&
  \bibinfo{author}{Andrei \surnamestart Voronkov\surnameend}
  (\bibinfo{year}{2002}): \emph{\bibinfo{title}{The Design and Implementation
  of VAMPIRE}}.
\newblock {\sl \bibinfo{journal}{AI Commun.}}
  \bibinfo{volume}{15}(\bibinfo{number}{2,3}), p. \bibinfo{pages}{91--110}.

\bibitemdeclare{inproceedings}{princess08}
\bibitem{princess08}
\bibinfo{author}{Philipp \surnamestart R{\"u}mmer\surnameend}
  (\bibinfo{year}{2008}): \emph{\bibinfo{title}{A Constraint Sequent Calculus
  for First-Order Logic with Linear Integer Arithmetic}}.
\newblock In: {\sl \bibinfo{booktitle}{Proceedings, 15th International
  Conference on Logic for Programming, Artificial Intelligence and Reasoning}},
  {\sl \bibinfo{series}{LNCS}} \bibinfo{volume}{5330},
  \bibinfo{publisher}{Springer}, pp. \bibinfo{pages}{274--289},
  \doi{10.1007/978-3-540-89439-1\_20}.

\bibitemdeclare{inproceedings}{starexec}
\bibitem{starexec}
\bibinfo{author}{Aaron \surnamestart Stump\surnameend}, \bibinfo{author}{Geoff
  \surnamestart Sutcliffe\surnameend} \& \bibinfo{author}{Cesare \surnamestart
  Tinelli\surnameend} (\bibinfo{year}{2014}): \emph{\bibinfo{title}{StarExec: A
  Cross-Community Infrastructure for Logic Solving}}.
\newblock In \bibinfo{editor}{St{\'e}phane \surnamestart Demri\surnameend},
  \bibinfo{editor}{Deepak \surnamestart Kapur\surnameend} \&
  \bibinfo{editor}{Christoph \surnamestart Weidenbach\surnameend}, editors:
  {\sl \bibinfo{booktitle}{Automated Reasoning}}, \bibinfo{publisher}{Springer
  International Publishing}, \bibinfo{address}{Cham}, pp.
  \bibinfo{pages}{367--373}, \doi{10.1007/978-3-319-08587-6\_28}.

\end{thebibliography}

\clearpage
\appendix
\clearpage
\section{Detailed results}
\label{app:detres} 

\begin{table}[ht]
\centering
\caption{Solver performance on LIA-lin track}
\begin{tabular}{lrrrrrr}
\hline
\textbf{Solver}& \textbf{Score} & \textbf{\#sat} & \textbf{\#unsat} & \textbf{CPU time/s} & \textbf{Wall-clock/s} & \textbf{\#unique}\\\hline\hline
\rowcolor{lightgray} \spacer & 338 & 235 & 103 & 299420 & 149835 & 36\\\hline
\golem & 309 & 215 & 94 & 374736 & 142604 & 25\\\hline
\eldarica & 307 & 219 & 88 & 372231 & 134933 & 38\\\hline
\unihorn & 169 & 107 & 62 & 551859 & 466284 & 0\\\hline
\treeautomizer & 139 & 81 & 58 & 633917 & 605367 & 0\\\hline
\end{tabular}
\label{tab:res_lia_lin}
\end{table}

\begin{table}[ht]
\centering
\caption{Solver performance on LIA-nonlin track}
\begin{tabular}{lrrrrrr}
\hline
\textbf{Solver}& \textbf{Score} & \textbf{\#sat} & \textbf{\#unsat} & \textbf{CPU time/s} & \textbf{Wall-clock/s} & \textbf{\#unique}\\\hline\hline
\rowcolor{lightgray} \spacer & 421 & 286 & 135 & 75414 & 39303 & 40\\\hline
\golem & 365 & 240 & 125 & 196890 & 196913 & 2\\\hline
\eldarica & 358 & 229 & 129 & 215589 & 76099 & 7\\\hline
\unihorn & 204 & 123 & 81 & 485808 & 391416 & 1\\\hline
\treeautomizer & 50 & 13 & 37 & 691499 & 648778 & 0\\\hline
\end{tabular}
\label{tab:res_lia_nonlin}
\end{table}

\begin{table}[ht]
\centering
\caption{Solver performance on LIA-lin-Arrays track}
\begin{tabular}{lrrrrrr}
\hline
\textbf{Solver}& \textbf{Score} & \textbf{\#sat} & \textbf{\#unsat} & \textbf{CPU time/s} & \textbf{Wall-clock/s} & \textbf{\#unique}\\\hline\hline
\rowcolor{lightgray} \spacer & 288 & 213 & 75 & 355686 & 178328 & 85\\\hline
\eldarica & 220 & 149 & 71 & 481857 & 163636 & 10\\\hline
\unihorn & 204 & 137 & 67 & 479998 & 393423 & 2\\\hline
\treeautomizer & 170 & 113 & 57 & 491989 & 470080 & 0\\\hline
\end{tabular}
\label{tab:res_lia_lin_arrays}
\end{table}

\begin{table}[ht]
\centering
\caption{Solver performance on LIA-nonlin-Arrays track}
\begin{tabular}{lrrrrrr}
\hline
\textbf{Solver}& \textbf{Score} & \textbf{\#sat} & \textbf{\#unsat} & \textbf{CPU time/s} & \textbf{Wall-clock/s} & \textbf{\#unique}\\\hline\hline
\rowcolor{lightgray} \spacer & 342 & 197 & 145 & 180474 & 95392 & 115\\\hline
\eldarica & 215 & 129 & 86 & 449249 & 177427 & 10\\\hline
\unihorn & 168 & 88 & 80 & 368162 & 297747 & 2\\\hline
\treeautomizer & 89 & 19 & 70 & 618917 & 488550 & 5\\\hline
\end{tabular}
\label{tab:res_lia_nonlin_arrays}
\end{table}

\begin{table}[ht]
\centering
\caption{Solver performance on LRA-TS track}
\begin{tabular}{lrrrrrr}
\hline
\textbf{Solver}& \textbf{Score} & \textbf{\#sat} & \textbf{\#unsat} & \textbf{CPU time/s} & \textbf{Wall-clock/s} & \textbf{\#unique}\\\hline\hline
\rowcolor{lightgray} \spacer & 317 & 234 & 83 & 355136 & 181996 & 37\\\hline
\golem & 311 & 235 & 76 & 364678 & 121607 & 19\\\hline
\treeautomizer & 155 & 114 & 41 & 646968 & 619460 & 4\\\hline
\unihorn & 103 & 65 & 38 & 725651 & 613022 & 1\\\hline
\end{tabular}
\label{tab:res_lra_ts}
\end{table}

\begin{table}[ht]
\centering
\caption{Solver performance on LRA-TS-parallel track}
\begin{tabular}{lrrrrrr}
\hline
\textbf{Solver}& \textbf{Score} & \textbf{\#sat} & \textbf{\#unsat} & \textbf{CPU time/s} & \textbf{Wall-clock/s} & \textbf{\#unique}\\\hline\hline
\rowcolor{lightgray} \spacer & 341 & 256 & 85 & 627685 & 319147 & 35\\\hline
\golem & 333 & 256 & 77 & 988621 & 329628 & 22\\\hline
\treeautomizer & 155 & 114 & 41 & 671801 & 641980 & 3\\\hline
\unihorn & 103 & 65 & 38 & 1072345 & 719996 & 1\\\hline
\end{tabular}
\label{tab:res_lra_ts_parallel}
\end{table}

\begin{table}[ht]
\centering
\caption{Solver performance on ADT-nonlin track}
\begin{tabular}{lrrrrrr}
\hline
\textbf{Solver}& \textbf{Score} & \textbf{\#sat} & \textbf{\#unsat} & \textbf{CPU time/s} & \textbf{Wall-clock/s} & \textbf{\#unique}\\\hline\hline
\ringen & 92 & 50 & 42 & 165240 & 163756 & 27\\\hline
\eldarica & 60 & 29 & 31 & 223335 & 73011 & 4\\\hline
\rowcolor{lightgray} \spacer & 55 & 25 & 30 & 226205 & 226257 & 5\\\hline
\treeautomizer & 0 & 0 & 0 & 1881 & 745 & 0\\\hline
\end{tabular}
\label{tab:adt_nonlin}
\end{table}

\begin{table}[ht]
\centering
\caption{Solver performance on LIA-nonlin-Arrays-nonrecADT track}
\begin{tabular}{lrrrrrr}
\hline
\textbf{Solver}& \textbf{Score} & \textbf{\#sat} & \textbf{\#unsat} & \textbf{CPU time/s} & \textbf{Wall-clock/s} & \textbf{\#unique}\\\hline\hline
\eldarica & 395 & 242 & 153 & 125994 & 46440 & 103\\\hline
\rowcolor{lightgray}\spacer & 298 & 179 & 119 & 131079 & 131124 & 6\\\hline
\treeautomizer & 0 & 0 & 0 & 4825 & 1894 & 0\\\hline
\end{tabular}
\label{tab:res_lia_nonlin_Arrays_nonrecADT}
\end{table}

\end{document}